\documentclass[aps,prl,superscriptaddress,amsfonts,amsmath,amssymb,showpacs,floatfix,reprint,longbibliography]{revtex4-2}

\usepackage{url}
\usepackage{bm}
\usepackage{graphicx}
\usepackage{epsfig}
\usepackage{amsmath}
\usepackage{amstext}
\usepackage{amssymb}
\usepackage{amsfonts}
\usepackage{amsbsy}
\usepackage{verbatim}
\usepackage{xcolor}
\usepackage{multirow}
\usepackage{floatrow}
\usepackage{float}
\usepackage{gensymb}
\usepackage{textcomp}
\usepackage{enumitem}
\usepackage[version=4]{mhchem}
\usepackage{afterpage}
\usepackage{pbox}
\usepackage{makecell}
\usepackage{array,booktabs}
\usepackage{dsfont}
\usepackage{siunitx}
\usepackage[utf8]{inputenc}
\usepackage{braket}
\usepackage[normalem]{ulem}

\graphicspath{ {FiguresMain/} }

\begin{document}

\title{Experimental and theoretical thermodynamic studies in Ba$_{2}$MgReO$_{6}$ - the ground state in the context of Jahn-Teller effect}

\author{Jana P\'{a}sztorov\'{a}*}
\email{jana.pasztorova@epfl.ch}
\affiliation{Laboratory for Quantum Magnetism, Institute of Physics, \'Ecole Polytechnique F\'ed\'erale de Lausanne, CH-1015 Lausanne, Switzerland}

\author{Aria Mansouri Tehrani}
\affiliation{Department of Materials, ETH Zurich, CH-8093 Zurich, Switzerland}

\author{Ivica {\v Z}ivkovi{\' c}}
\affiliation{Laboratory for Quantum Magnetism, Institute of Physics, \'Ecole Polytechnique F\'ed\'erale de Lausanne, CH-1015 Lausanne, Switzerland}

\author{Nicola A. Spaldin}
\affiliation{Department of Materials, ETH Zurich, CH-8093 Zurich, Switzerland}

\author{Henrik M. Rønnow}
\affiliation{Laboratory for Quantum Magnetism, Institute of Physics, \'Ecole Polytechnique F\'ed\'erale de Lausanne, CH-1015 Lausanne, Switzerland}

\date{\today}



\begin{abstract}

{We address the degeneracy of the ground state multiplet on the 5$d^1$ Re$^{6+}$ ion in double perovskite Ba$_{2}$MgReO$_{6}$ using a combination of specific heat measurements and density functional calculations. For Ba$_{2}$MgReO$_{6}$, two different ground state multiplets have previously been proposed - a quartet (with degeneracy $N$=4) \cite{jr:hirai} and a doublet ($N$=2) \cite{jr:marjerrison}. Here we employ two independent methods for the estimation of phonon contribution in heat capacity data to obtain the magnetic entropy $S_{mag}$, which reflects the degeneracy of the ground state multiplet $N$ through $S_{mag}=R$ln$N$. In both cases, we obtain a better fit to $S_{mag}=R$ln2 indicating evidence of $N$=2 degeneracy in the range from 2 to 120~K. The detailed nature of the ground state multiplet in Ba$_{2}$MgReO$_{6}$ remains an open question.}
  
\end{abstract}

\maketitle

\section{Introduction}
	 
In recent years there has been an increase in interest in materials which exhibit strong spin-orbit coupling (SOC) with both experimental and theoretical studies revealing exotic phases, such as spin-liquids \cite{jr:spin_liquid,jr:greedan_spinL} or systems with multipolar orders \cite{jr:chen,jr:exo-balents,jr:gaulin}. Since the strength of SOC grows rapidly with increasing atomic number $Z$, it is important in materials containing 5\textit{d} ions, where the entanglement between spin and orbital degrees of freedom profoundly affects the low-energy excitation spectrum.

\begin{figure}[t]
 	\includegraphics[width=0.7\columnwidth]{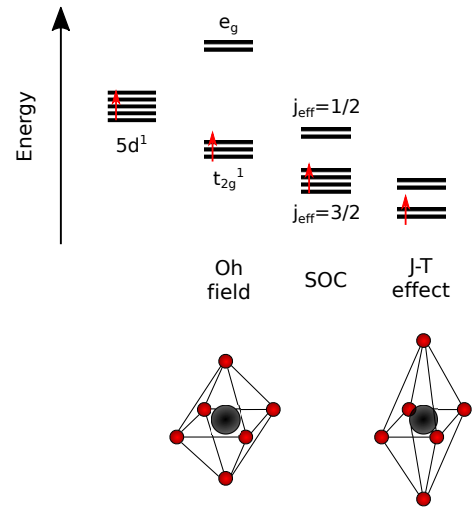}
 	\centering
 	\caption{Proposed energy level splitting scheme for 5$d^{1}$ electronic configuration in DPs. Elongation of ideal Re-O$_{6}$ octahedron along $c$-axis \cite{jr:hirai_rex}, known as Jahn-Teller (J-T) effect, lifts degeneracy of $d$ orbitals and splits original $j_{eff}$=3/2 quartet into two Kramer doublets. The splitting of higher energy levels is not shown for clarity.} 
 	\label{fig:energy}
 \end{figure}

One such family of compounds comprises double perovskites containing 5\textit{d} elements Ta, Re and Os with formally single electron occupying their \textit{d} orbitals (5$d^1$). These have general chemical formula $A_{2}BB'O_{6}$, with the magnetic $B'$ ion surrounded by six anions, creating a highly symmetric octahedron. The octahedral crystal electronic field (CEF) splits the five-fold degenerate \textit{d} orbitals into characteristic $e_{g}$ and $t_{2g}$ orbital sets (see Fig.~\ref{fig:energy}). Under the influence of SOC the three-fold degenerate $t_{2g}$ set (with an effective orbital momentum $l_{eff} = -1$) is split further into a lower-lying quartet ($j_{eff}$=3/2) and a doublet ($j_{eff}$=1/2) at higher energy. Up to this point, the single ion ground state multiplet is the quartet ($N$=4) proposed in \cite{jr:hirai} paper. However, further investigation of some double perovskites $A_{2}BB'O_{6}$ showed the structural distortion from cubic to tetragonal unit cell and elongation or compression of the ideal $B'$-O$_{6}$ octahedon along $c$-axis \cite{jr:hirai_rex,jr:ishikawa}. In theory, this is known as Jahn-Teller effect that leads to another splitting of ground quartet into two Kramer doublets, where $N$=2, and which translates into $Rln$~2 value of entropy.

The coupling between magnetic moments has been considered to arise from a \textit{super-} superexchange mechanism, comprising both feromagnetic ($J_{FM}$) and antiferromagnetic ($J_{AFM}$) type and leading to a long-range magnetic order at $T_m$~\cite{jr:chen}. Additionally, it has been suggested that 5$d^1$ systems could allow investigation of higher-order multipoles due to their small dipole magnetic moments~\cite{jr:chen}. For an electronic quadrupolar interaction $V$ that is large enough, the prediction of a quadrupolar order at $T_q > T_m$ has been made \cite{jr:chen}. This causes a small lattice distortion from a perfect cubic environment.

Several 5$d^1$ double perovskites have been previously investigated in this context. In Ba$_{2}$NaOsO$_{6}$, where the Os ion is in a formally 7+ valence state, a long-range magnetic ordering has been found to occur around $T_{m}$~=~7 K~\cite{jr:osmiumCG,Steele2011,jr:Os_nat_com,jr:Os_2shoulders}, preceded by a breaking of local point symmetry, indicated by the splitting of NMR spectra below 12 K~\cite{jr:Os_nat_com}. Specific heat studies revealed a sharp transition at $T_m$~\cite{jr:osmiumDP,jr:Os_2shoulders} and a small, broad shoulder extending towards higher temperatures. Authors assigned this shoulder to the nematic (orthogonal paramagnetic) phase based on temperature derivative of specific heat data ~\cite{jr:Os_2shoulders}. For $B$ = Li only a single transition has been observed~\cite{jr:osmiumCG,Steele2011} and no detailed specific heat studies have been reported. In the rhenium family several compounds have been investigated, all with Re in a formally 6+ valence state and the $B$ site occupied by 2+ ions. For $B$ = Mg, $T_m$ occurs at 18 K~\cite{jr:marjerrison,jr:hirai}, with a broad shoulder revealed in calorimetric data around 33 K. This shoulder has been associated with a cubic to tetragonal structural distortion and argued to indicate the quadrupolar order ~\cite{jr:hirai_rex}. DFT calculations further demonstrated the ordering of the proposed charge quadrupoles and suggested an additional spin-canting dependent ordered charge quadrupolar component\cite{mansouri_tehrani_untangling_2021}. A similar sequence of features has been revealed in Ba$_{2}$ZnReO$_{6}$ ($T_m = 16$ K, $T_q = 23$ K)~\cite{Barbosa2020} and Ba$_{2}$CdReO$_{6}$ ($T_m = 4$ K, $T_q = 25$ K)~\cite{jr:hirai_Cd}, pointing to a universal behavior across the family. Additionally, charge quadrupoles have also been recently discussed in the more ionic vacancy-ordered 5$d^1$ double perovskite halides such as in Cs$_2$TaCl$_6$\cite{tehrani_charge_2022}.

Coming back to earlier discussion about the ground state multiplet of 5$d^{1}$ systems, when the measured specific heat is used to extract the magnetic entropy, it has been observed in several reports to amount to a value significantly lower than the expected $R$ln4~\cite{jr:osmiumDP,jr:marjerrison,Barbosa2020}. For Ba$_{2}$NaOsO$_{6}$ it has been suggested that the missing entropy should be recovered at higher temperatures~\cite{jr:osmiumDP}, although the NMR spectra show that the local environment is not distorted above $\sim 12$ K~\cite{jr:Os_nat_com}. In contrast, recent single crystal experiments on Ba$_{2}$MgReO$_{6}$~\cite{jr:hirai} indicate a complete $R$ln4 recovery, with the magnetic entropy being released all the way up to 80 K, much higher than $T_q = 33$ K. Similarly to the case of Ba$_{2}$NaOsO$_{6}$, an ideal cubic environment has been found in high resolution synchrotron diffraction experiment at $T > T_q = 33$ K~\cite{jr:hirai_rex}. Thus, there still remains an open question of the relationship between the magnetic entropy, ordering and ground state in 5$d^1$ double perovskites.

In this article we focus our attention on the phonon background in Ba$_{2}$MgReO$_{6}$, which is a crucial component for obtaining the magnetic specific heat. There are several ways to estimate the phonon contribution including analytical approaches, first-principle calculations and by measuring a nonmagnetic analog in which the magnetic ion has been replaced with a non magnetic one. Ref.~\cite{jr:hirai} used an analytical approach and showed that the sum of two Debye functions could describe the data above 80 K, although, there was no further clarification for use of these two Debye temperatures (299 K and 796 K) in particular, nor was the optical contribution addressed through Einstein modes. In this work, we determine of the phonon contribution by comparing measurments on newly synthesized powders of Ba$_{2}$MgReO$_{6}$ and its nonmagnetic analog Ba$_{2}$MgWO$_{6}$. We complement our measurements with density-functional calculations of the phonon density of states, from which we extract the phonon specific heat.

\section{Methods}
	 
\subsection{Experimental Details}

Polycrystalline samples of Ba$_{2}$MgReO$_{6}$ and Ba$_{2}$MgWO$_{6}$ were prepared by a standard solid-state method following published papers \cite{jr:WO6_pow,jr:ReO6_pow}. For Ba$_{2}$MgReO$_{6}$, stoichiometric amounts of BaO, MgO and ReO$_{3}$ were mixed and ground in an agate mortar under a protective argon atmosphere in a glove box. The mixture was placed into silica tubes, evacuated and sealed. The tubes were annealed in a box furnace for 24 hours at 900 $^{\circ}$C and cooled at the furnace cooling rate. For Ba$_{2}$MgWO$_{6}$, the polycrystalline powder was prepared by repeated annealing and intermediate grinding. Powders of BaCO$_{3}$, MgO and WO$_{3}$ were mixed in 2:1:1 ratios and ground in an agate mortar. The mixture was compressed into the pellet and heated up to 850 $^{\circ}$C for 24 hours. The initial reaction was followed by two annealing cycles with intermediate grinding and compression into a new pellet. First annealing was done at 1150 $^{\circ}$C for 20 hours and the second one at 1150 $^{\circ}$C for 12 hours.

The chemical composition of the polycrystalline samples was determined by powder X-ray diffraction on a Panalytical  Empyrean diffractometer equipped with a PIXcel-1D detector, Bragg-Brentano beam optics and utilising a monochromated Cu K$\alpha_{1,2}$ source ($\lambda_{1}$~=~1.5406~{\AA}) at 300 K. The refinements of all samples using the \textit{HighScore Plus} software were consistent with previous reports on powder samples \cite{jr:ReO6_pow,jr:WO6_pow}.

Specific heat measurements were performed using the relaxation method in a commercial apparatus (PPMS, Quantum Design). All polycrystalline samples were compressed into pellets with diameter 3.2 mm, thickness under 0.5 mm and mass $\sim$ 10 mg.

\subsection{Computational Details}

Density functional theory (DFT) calculations were performed using a plane wave basis set and projected augmented wave (PAW) potentials as implemented in the Vienna $ab-initio$ simulation package (VASP)\cite{kresse_ab_1993, kresse_efficient_1996, kresse_ultrasoft_1999, blochl_projector_1994}. We used the following pseudopotentials: Ba$_{sv}$ ($5s^2\,5p^6\,6s^2$), Mg$_{pv}$ ($2p^6\,3s^2$), Re$_{pv}$ ($5p^6\,5d^5\,6s^2$), and O ($2s^2\,2p^4$).
The Perdew-Burke-Ernzerhof (PBE) generalized gradient approximation was utilized to account for exchange and correlation, with an effective on-site Hubbard U = 1.8\,eV correction applied to the $d$ orbitals of the magnetic ions within the Dudarev scheme\cite{dudarev_electron-energy-loss_1998}. In addition, we included spin-orbit coupling using a full relativistic scheme\cite{steiner_calculation_2016}. 
An energy cutoff of 600\,eV and a gamma-centered $k$-point mesh of $6\times6\times6$ were used. Electronic and structural optimizations were carried out with energy convergence criteria of $1\times10^{-6}$\,eV. The crystal structure including the internal atomic positions were relaxed, while the experimental $c/a$ ratio was constrained. The relaxed lattice parameters are $a = 5.8$~{\AA} and $c = 8.2$~{\AA}. The magnetic structure, canted in-plane antiferromagnetism, was imposed as described in Ref.~\onlinecite{mansouri_tehrani_untangling_2021} by constraining the magnetic moments on the Re sites. The vibrational properties of Ba$_{2}$MgReO$_{6}$ were calculated using the open-source PHONOPY package\cite{togo_first_2015}, which constructs a force-constant matrix (calculated by VASP) by displacing symmetry-independent atoms by $\pm{0.01}\AA$ from their equilibrium positions. Since DFT is a zero kelvin technique, phonon and therefore specific heat calculations are only calculated for the tetragonal ground-state crystal structure.

\section{Experimental Results}

\subsection{Phonon contribution to specific heat using the non-magnetic analogue}

\begin{figure}[t]
	\includegraphics[width=0.8\columnwidth]{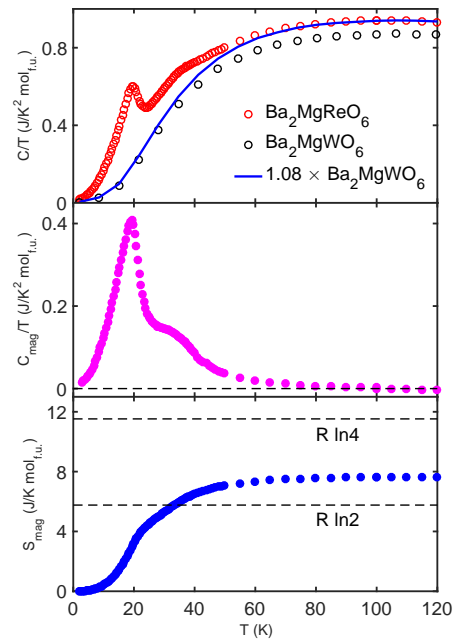}
	\centering
	\caption{(a) Temperature dependence of specific heat divided by temperature $C/T$ measured for Ba$_2$MgReO$_6$ powder (5$d^1$), red circles, and Ba$_2$MgWO$_6$ powder (5$d^0$), black circles. The solid blue line represents Ba$_2$MgWO$_6$ data scaled by 1.08. (b) Difference $C_{mag}/T$ between magnetic and non-magnetic sample. (c) Entropy plot obtained from integrated $C_{mag}/T$ for the Ba$_2$MgReO$_6$ powder.} 
	\label{fig:nonmag}
\end{figure}

In Fig.~\ref{fig:nonmag}(a) we present the measured specific heat of Ba$_{2}$MgReO$_{6}$ and Ba$_{2}$MgWO$_{6}$ powders. The Re compound exhibits two features reported in previous publications, a sharp peak at $T_m = 18$ K and a broad feature around $T_q = 33$ K. On the other hand the W compound displays a smooth behavior across the investigated range. As Ba$_{2}$MgReO$_{6}$ and Ba$_{2}$MgWO$_{6}$ are isostructural compounds, it is reasonable to expect similar phonon contribution of the specific heat as a function of temperature. In order to use the Ba$_{2}$MgWO$_{6}$ specific heat as estimate of the phonon specific heat in the Ba$_{2}$MgReO$_{6}$ compound, the  W amplitude was scaled by a factor of 1.08 to match the high temperature part of the Re specific heat. There is also a small shift between the two curves along the temperature axis that is the result of different atomic masses of the two compounds, causing dissimilar characteristic Debye temperatures. We dismiss errors in heat capacity measurements that are lower than 0.5$\%$. Minimal amount of low-temperature grease was used with negligible contribution to data inaccuracy.

We subtract this renormalized W curve from the Re data, presented in Fig.~\ref{fig:nonmag}(b). The magnetic ordering at $T_m$ and the broad bump around $T_q$ are now emphasized, and the tail of the magnetic contribution can be seen to diminish above $\sim 60$ K. The magnetic entropy is then obtained by integrating $S_{mag} = \int (C_{mag}/T) dT$, and the result is presented in Fig.~\ref{fig:nonmag}(c). The saturation value approaches 8 J/Kmol, which is significantly lower than the expected $R$ln4 = 11.5 J/Kmol.

\begin{figure}[h!]
	\includegraphics[width=60mm]{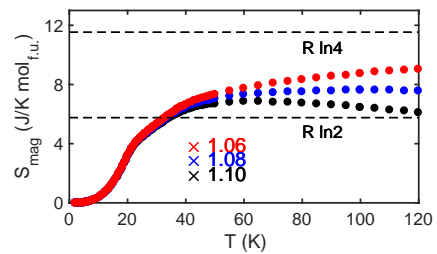}
	\centering
	\caption{Entropy plot obtained from integrated $C_{mag}/T$ using phonon background of non-magnetic Ba$_2$MgWO$_6$ studied in limits of different normalization factors - from 1.06 to 1.10. Negative slope or rapid increase of entropy in region above 60~K does not have physical meaning. The normalization factor of 1.08 was used as most suitable in Fig.~\ref{fig:nonmag}.}
	\label{fig:sc_f_W}
\end{figure}

In Fig.~\ref{fig:sc_f_W} we plot the magnetic entropy as a function of temperature using different scaling factors for phonon part of specific heat to observe which values do not have physical meaning. The value of 1.08 is chosen in order to have the high temperature entropy approximately saturated, implying that the magnetic degrees of freedom are engaged only at low temperatures.

\subsection{Phonon contribution to specific heat using first principles calculations}

 \begin{figure}[t]
	\includegraphics[width=0.8\columnwidth]{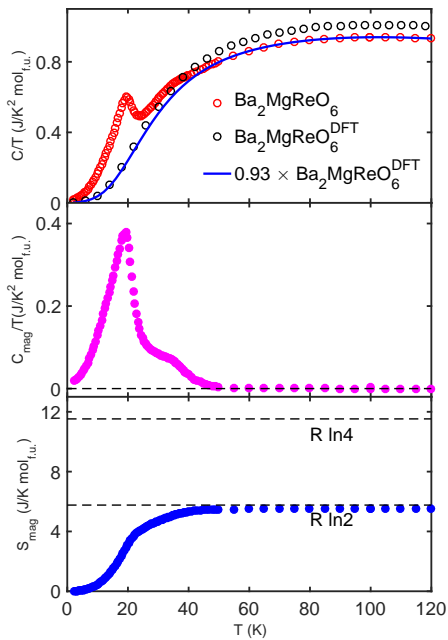}
	\centering
	\caption{(a) Temperature dependence of specific heat divided by temperature $C/T$ - experimental data for Ba$_2$MgReO$_6$ (red) and the phonon contribution for Ba$_2$MgReO$_6$ calculated from DFT (black). Solid blue line represents scaled DFT data. (b) $C_{mag}/T$ shows the difference between experimental Ba$_2$MgReO$_6$ and DFT data. (c) Entropy plot obtained from integrating $C_{mag}/T$. The differences between two methods in Fig. \ref{fig:nonmag} and \ref{fig:DFT} are very subtle - in case of non-magnetic analogue method, the second anomaly in $C_{m}/T$ data is bigger and decreases to zero values slower.} 
	\label{fig:DFT}
\end{figure}

The calculated phonon specific heat is shown in Fig.~\ref{fig:DFT}(a), black circles, and presents similar temperature dependence as data. As with the non-magnetic analog, we observe a slight mismatch between the maxima in $C/T$, in this case arising from two possible factors - one being an error in the calculation of interatomic forces and consequently the dynamic matrix, another is related to the approximation in phonon calculations (in case of DFT, the interatomic forces are calculated at zero kelvin and then phonons are elucidated using a harmonic approximation). Therefore we employ a renormalization factor to match the high temperature experimental data of Ba$_{2}$MgReO$_{6}$. 

Fig.~\ref{fig:DFT}(b) displays the subtraction of the experimental data and the calculated phonon background, revealing a similar shape as the $C_{mag}/T$ in Fig.~\ref{fig:nonmag}(b), with small differences above the magnetic transition - the non-magnetic analogue approach results in a slightly stronger $T_{q}$ peak and a tail decaying slower towards high temperature. Above 50~K, the magnetic contribution $C_{mag}/T$ oscillates very closely around $C_{mag}/T$=0 value and start to increase slightly at 130~K meaning that the fit of high temperature data is not perfect (the deviation is less than 3$\%$ at 200~K). We attribute this mismatch to the discrepancies between $C/T$ maxima of DFT and experimental data at higher temperatures. The resulting entropy in the range up to 120~K, Fig.~\ref{fig:DFT}(c), saturates above 50~K to a value within 95$\%$ of $R$ln2.

 
\section{Discussion}

The difference in extracted $S_{mag}$ utilising two methods in our work illustrates that the determination of the intrinsic phonon background of Ba$_{2}$MgReO$_{6}$ is not a straight forward process. Both the non-magnetic analog Ba$_{2}$MgWO$_{6}$ and the DFT calculations have their maximum in $C/T$ slightly shifted towards higher temperatures compared to Ba$_{2}$MgReO$_{6}$. This results in an imperfect subtraction, causing $C_{mag}/T$ values to oscillate around $C_{mag}/T$=0 for temperatures above 120~K. Nevertheless, it is evident from the extracted results that the magnetic entropy up to $\sim 120$ K is significantly below the expected \textit{R}ln4 for N=4.

Our results are in general agreement with several other publications where magnetic entropy was extracted in 5$d^{1}$ systems, covering both Os~\cite{jr:osmiumDP} and Re compounds~\cite{jr:marjerrison,Ishikawa2021}. This indicates that the choice of the phonon background presented in Ref.~\cite{jr:hirai} should not be regarded as a definitive proof of the total magnetic entropy reaching \textit{R}ln4. Another point to make is the presence of a strong SOC for this family of compounds and structural distortions of ideal octahedron present in these compounds at $T_{q}$ that might cause further splitting of single ion ground state doublet into two Kramer doublets. As more 5$d^1$ systems are emerging, it opens possibilities to study this ground state multiplet further, even employing more powerful experimental methods. The determination of the phonon part of heat capacity is non-trivial and require careful consideration.

\section{Summary}

We determined the phonon contribution to the heat capacity of Ba$_2$MgReO$_6$ by measuring its non-magnetic analog Ba$_2$MgWO$_6$ as well as calculating it from first principles. We find that both approaches give a slightly mismatched maximum in $C/T$, indicating a non-trivial problem of establishing phonon contribution at high temperatures. Both approaches result in total magnetic entropy reaching values significantly lower than $R$ln4, which will hopefully instigate more detailed investigations to resolve this controversial issue.

	 \section*{Acknowledgements}
	 This work was funded by the European Research Council (ERC) under the European Union’s Horizon 2020 research and innovation program projects HERO (Grant No. 810451). Jana P\'{a}sztorov\'{a} acknowledges support from the Federal Commission for Scholarships for Foreign Students for the Swiss Government Excellence Scholarship (ESKAS No. 2019.0041) for the academic year 2019-20.

 \bibliographystyle{unsrt}

 \bibliography{journals,books,Aria}

\end{document}